\documentstyle[preprint,prl,aps]{revtex}

\begin{document}

\draft

\preprint{\vbox{
\hbox{CTP-TAMU-01/95}
\hbox{hep-ph/9501247}
\hbox{January 1995}
}}

\title{Planck Scale Physics and the Testability of SU(5) Supergravity GUT}

\author{D.~Ring\cite{dave}, S.~Urano\cite{shinichi}, and
R.~Arnowitt\cite{dick}}

\address{Center for Theoretical Physics, Department of Physics,
		Texas A\&M University,
  		College Station, TX 77843-4242}

\maketitle

\begin{abstract}

GUT scale threshold corrections in minimal SU(5) supergravity grand unification
are discussed.  It is shown that predictions may be made despite uncertainties
associated with the high energy scale. A bound relating the strong coupling
constant to the mass scales associated with proton decay and supersymmetry is
derived, and a sensitive probe of the underlying theory is outlined.  In
particular, low energy measurements can in principle determine the presence of
Planck scale ($ 1 / {{\rm M}_{\rm Pl}} $) terms.

\end{abstract}

Over the past two decades, much attention has been given to the possibility of
unifying the three gauge groups of the Standard Model into one group. These GUT
models are theoretically more appealing than the Standard Model for various
reasons \cite{arnowitt}.  The 1990 precision LEP data strongly indicated that
supersymmetry is needed to achieve grand unification \cite{amaldi_etc}, and
spurred on many new analyses of different aspects of SUSY GUTs.

With the increasing precision of data, further predictions will be affected by
supersymmetry threshold effects and the details of the model at high energy. In
this note we examine the effects of the non-degeneracy of the super-heavy GUT
spectrum, and the possible existence of non-renormalizable operators from
Planck scale physics. A number of treatments of these issues have been given
recently for the SU(5)
model\cite{barbieri/hall,faraggi,hall/sarid,langacker/polonsky}. It was found
that many of the predictions for low energy observables were blurred by the
high scale effects.  For example, it was argued that the SUSY scale cannot be
determined by a more accurate measurement of $\alpha_3$ \cite{barbieri/hall},
and the rate of proton decay cannot be predicted from low energy data if
additional Planck scale terms are present \cite{hall/sarid}.

Nevertheless, we will show here that there are still predictions to be made in
this model:

\paragraph{} The effect of Planck scale non-renormalizable terms becomes
smaller as the value of $\alpha_3$ is varied to lower values, so the lower
limit on $\alpha_3$ is not lost.  This is especially interesting since there is
currently a disparity in the values of $\alpha_3$ between the measurements made
at weak scale energies, and those made at lower energies \cite{foot1}.
Resolving this disparity and refining the measured value of $\alpha_3$ will
provide an important test of the model.

\paragraph{} Since Planck scale physics smears the correlation between
$\alpha_3$ and the mass scale which governs proton decay, proton decay will be
a sensitive probe of Planck scale physics.  In particular it will be seen that
by purely low-energy measurements one can determine experimentally the degree
to which the dominant Planck scale term is present.  Thus models of this type
allow for the first time a test for the existence of Planck scale physics and
whether Planck scale physics impinges on low-energy (electroweak scale)
physics.

With respect to the second item, progress has been made recently in deriving
models similar to the ones considered here from string
theory\cite{adjoint,rulin}, so that the gravitational smearing may be
calculable in principle. Proton decay would then be a sensitive test of string
theory.

Our model of GUT physics is defined by superpotential\cite{su5gut,su5gutmore}
\begin{eqnarray}
W &=& \lambda_1 [ {\scriptstyle{1\over3}} {\, \rm tr}(\Sigma^3) +
                       {\scriptstyle{1\over2}} {\rm M} {\, \rm tr}(\Sigma^2) ]
		\nonumber \\
  &&+ \lambda_2 \overline{{\rm H}}_X ( \Sigma^X_{\ Y}
	+ 3 {\rm M}' \, \delta^X_{\ Y}) {\rm H}^Y
			\nonumber \\
  &&+ \varepsilon_{UVWXY} \, {\rm H}^U {\rm M}^{VW} f_1 {\rm M}^{XY} +
        \overline{{\rm H}}_X {\rm M}^{XY} f_2 \overline{{\rm M}}_Y,
\end{eqnarray}
where $\Sigma, {\rm H},$ and $\overline{{\rm H}}$ form a {\bf 24}, {\bf 5}, and
{$\bf \overline{5}$} of SU(5) respectively, $\overline{{\rm M}}$ and ${\rm M}$
are {$\bf \overline{5}$} and {\bf 10} matter superfields, $f_1$ and $f_2$ are
Yukawa coupling constant matrices in the generation space, and the mass
parameters ${\rm M}$ and ${\rm M}'$ are set equal to account for a light Higgs
doublet.  (This is a well known fine tuning problem with this model.  We
consider elsewhere alternate models such as those in Ref.~\cite{su6model} which
avoid this fine tuning.)  The gauge group is broken down to SU(3) $\times$
SU(2) $\times$ U(1) when $\Sigma$ grows a VEV: $ \langle \Sigma \rangle = {\rm
M} {\, \rm diag}(2,2,2,-3,-3)$. The resulting super-heavy spectrum includes a
heavy color Higgs chiral multiplet ({\bf 3},{\bf 1},$2\over3$) of mass ${\rm
M}_{\rm H}=5 \lambda_2 {\rm M}$, a vector multiplet ({\bf 3},{\bf 2},$5\over3$)
of mass ${\rm M}_{\rm V} = 5 \sqrt{2} g {\rm M}$, chiral multiplets ({\bf
8},{\bf 1},0) and ({\bf 1},{\bf 3},0) of mass ${\rm M}_\Sigma = {5\over2}
\lambda_1 {\rm M}$, and a Standard Model gauge singlet chiral multiplet ({\bf
1},{\bf 1},0) of mass $\frac{1}{2} \lambda_1 {\rm M}$, where the numbers in
parentheses are the SU(3) and SU(2) representations and hypercharge quantum
numbers.

In the following we assume $0.1 \le \lambda_{1,2} \le 2.0$, i.e. $10^{-3}
\lesssim \alpha_{\lambda_{1,2}} \lesssim 1/3$.  The upper bound is imposed so
that the model stays within the perturbative domain, while the lower bound
excludes any anomalously small couplings.

In addition to the renormalizable interactions, one may add the dominant
non-renormalizable operator from Planck scale physics \cite{hill_etc},
\begin{equation}
{\cal L}_0 = {c\over2{{\rm M}_{\rm Pl}}} {\, \rm tr}({\rm F} {\rm F} \Sigma),
\label{eq_pslop}
\end{equation}
where ${{\rm M}_{\rm Pl}} = 1/\sqrt{\kappa} = 1/\sqrt{8 \pi G}$.  In the first
part of this paper we impose, for naturalness, $|c|<1$.  The main effect of
this term is to modify the unification condition when $\Sigma$ grows a VEV.
Note for now that $\langle \Sigma \rangle$ will be $O({{\rm M}_{\rm GUT}})$ so
this term enters with a coefficient ${{{\rm M}_{\rm GUT}} / {{\rm M}_{\rm Pl}}}
\approx (1/10 - 1/100)$ and we would naively expect its effects to be small
\cite{foot2}.

We will concentrate mainly on gauge coupling unification. The running of the
gauge couplings with respect to the energy scale, $\mu$, is governed by the
two-loop renormalization group equations\cite{foot7},
\begin{equation}
	{d\over{dt}}{1\over{\alpha_i(t)}} =
	- b_i
	- {1\over{4 \pi}} \sum_{j} b_{ij} \, \alpha_j(t)
	+ \frac{1}{16 \pi^2} \sum_{f} b_{if} \, h_f^2(t)
\label{eq_run}
\end{equation}
where $\alpha_i \equiv { g_i^2 / {4 \pi}}$, $t \equiv (\ln \hat \mu) / {2
\pi}$, $\hat \mu = \mu/$(arbitrary mass parameter), and $h_f$ are the Yukawa
couplings.  In MSSM, one-loop coefficients are $b_i = (33/5,1,-3)_i$.  The
two-loop coefficients are also well known and can be found elsewhere, e.g.\ in
\cite{einhorn/jones}.

We first discuss the effect of GUT scale thresholds without the Planck term.
The GUT degrees of freedom are included in the running at their respective
thresholds by
\begin{equation}
{\alpha_i}^{-1}(\mu) = {\alpha_{i0}}^{-1}(\mu) - \sum_a \Delta b_{ia}^{th}
	{1\over{2 \pi}} \ln({\mu\over{\rm M}_{a}}),
\label{eq_alpha}
\end{equation}
where $\alpha_{i0}(\mu)$'s are calculated numerically to two-loop accuracy from
their low energy values via the RGE's using the MSSM beta functions. The low
energy values of $\alpha_{i0}(\mu)$'s include SUSY threshold which will be
discussed later. In Eq.~(\ref{eq_alpha}), the index $a$ sums over the GUT
degrees of freedom with masses less than $\mu$, $\Delta b_{i \Sigma}^{th} =
(0,2,3)_i $, $\Delta b_{i {\rm H}}^{th} = (2/5,0,1)_i $, and $\Delta b_{i {\rm
V}}^{th} = (-10,-6,-4)_i$. The largest of the ${\rm M}_a$ is called ${\rm
M}_{\rm U}$, as this is where the coupling constants actually meet. Thus the
unification condition is $\alpha_i({\rm M}_{\rm U}) = \alpha_5({\rm M}_{\rm
U})$.

On the other hand, if the Planck term is included, then when the VEV of
$\Sigma$ is inserted into the dominant Planck scale operator of
Eq.~(\ref{eq_pslop}), the kinetic terms for the gauge bosons will receive a
contribution. Thus the unification condition will be modified by replacing
$\alpha_5({\rm M}_{\rm U})^{-1}$ by
\begin{equation}
\alpha_5^{-1}({\rm M}_{\rm U}) (1 - c \, {{\rm M} \over {{\rm M}_{\rm Pl}}},
		1 - 3 c\, {{\rm M} \over {{\rm M}_{\rm Pl}}},
		1 + 2 c \, {{\rm M} \over {{\rm M}_{\rm Pl}}}),
\label{eq_mod}
\end{equation}
where ${\rm M}$ is the mass parameter entering in $\langle \Sigma \rangle$.

At low energies we must consider the decoupling of supersymmetric degrees of
freedom. This can be described at the one-loop level by three SUSY threshold
parameters ${\rm M}_i$, one for each coupling constant
\cite{langacker/polonsky}. The meaning of these parameters is as follows: if
above ${\rm M}_i$ we assume the threshold particles to be massless, and below
${\rm M}_i$ we assume them to be completely integrated out, and we assume the
couplings meet smoothly at ${\rm M}_i$, then at scales far from ${\rm M}_i$ our
running coupling constants will match the exact ones \cite{foot3}. Such an
${\rm M}_i$ can always be found for each $i$ so long as the one-loop beta
function above the threshold is different from that below \cite{foot4}. Thus
the effect of SUSY thresholds is given by \cite{foot5}
\begin{equation}
{\alpha_i}^{-1}({\rm M}_{\rm Z})
	= {\alpha_{i0}}^{-1}({\rm M}_{\rm Z}) + \Delta b_{i}^{th}
	  {1\over{2 \pi}} \ln({{\rm M}_{\rm Z}\over{\rm M}_{i}}).
\label{eq_threshold}
\end{equation}
Here, $\alpha_i({\rm M}_{\rm Z})$ are the couplings at ${\rm M}_{\rm Z}$, while
$\alpha_{i0}({\rm M}_{\rm Z})$ are the couplings one would obtain at ${\rm
M}_{\rm Z}$ if one ran with the full SUSY beta function down to ${\rm M}_{\rm
Z}$. $\Delta b_i^{th}=(5/2,25/6,4)_i$ gives the contribution to the $\beta$
function from the additional SUSY degrees of freedom. This is sufficient for a
two-loop analysis as well, so long as the SUSY thresholds are not too far from
${\rm M}_{\rm Z}$. We treat all Standard Model degrees of freedom except the
top as degenerate with or lighter than ${\rm M}_{\rm Z}$.  We take the top mass
to be $174 \ {\rm GeV}$ in accordance to the latest experimental indication
\cite{cdf}.  Uncertainties in its value do not affect our results
significantly.

There are two subtleties involved in relating the ${\rm M}_i$'s to the
sparticle spectrum. First, the coupling constants do not actually jump in slope
when the scale reaches the mass of a particle, rather they change gradually. If
our measurements are performed in the region of changing slope, ``match and
run'' will be inaccurate. Second, if the threshold region is close to the
electroweak symmetry breaking scale, there are always mass splittings among the
particles in the gauge multiplets. This potentially large effect has never been
fully treated, and is essential for incorporating detailed SUSY spectra in
unification analyses. Regardless of these subtleties, any supersymmetric
spectrum can be accommodated by Eq.~(\ref{eq_threshold}) if the ${\rm M}_i$'s
are allowed to vary below ${\rm M}_{\rm Z}$ as well as above
\cite{langacker/polonsky}.

Combining Eqs.~(\ref{eq_run}), (\ref{eq_alpha}), (\ref{eq_threshold}), and the
unification condition as modified by (\ref{eq_mod}), we arrive at the equation
which we use for our calculations:
\begin{eqnarray}
\lefteqn{
\alpha_5^{-1}({\rm M}_{\rm U})
\left( 1 - c \, {{\rm M} \over {{\rm M}_{\rm Pl}}},
	1 - 3 c\, {{\rm M} \over {{\rm M}_{\rm Pl}}},
	1 + 2 c \, {{\rm M} \over {{\rm M}_{\rm Pl}}}\right)_i  }
		\nonumber \\
&=& \alpha_i^{-1}({\rm M}_{\rm Z})
       - b_i \frac{1}{2 \pi} \ln(\frac{{\rm M}_{\rm U}}{{\rm M}_{\rm Z}})
		\nonumber \\
&&- \sum_j b_{ij} \frac{1}{4 \pi} \int_{t_{\rm Z}}^{t_{\rm U}} \! \!
		\alpha_{j 0}(t) \, dt
    + \sum_f b_{if} \frac{1}{16 \pi^2} \int_{t_{\rm Z}}^{t_{\rm U}} \! \!
                h_f^2(t) \, dt
		\nonumber \\
&&- \Delta b_i^{th} \frac{1}{2 \pi} \ln(\frac{{\rm M}_{\rm Z}}{{\rm M}_i})
    - \sum_a \Delta b_{ia}^{th} \frac{1}{2 \pi}
		\ln(\frac{{\rm M}_{\rm U}}{{\rm M}_a}).
\label{eq_main}
\end{eqnarray}

\begin{figure}
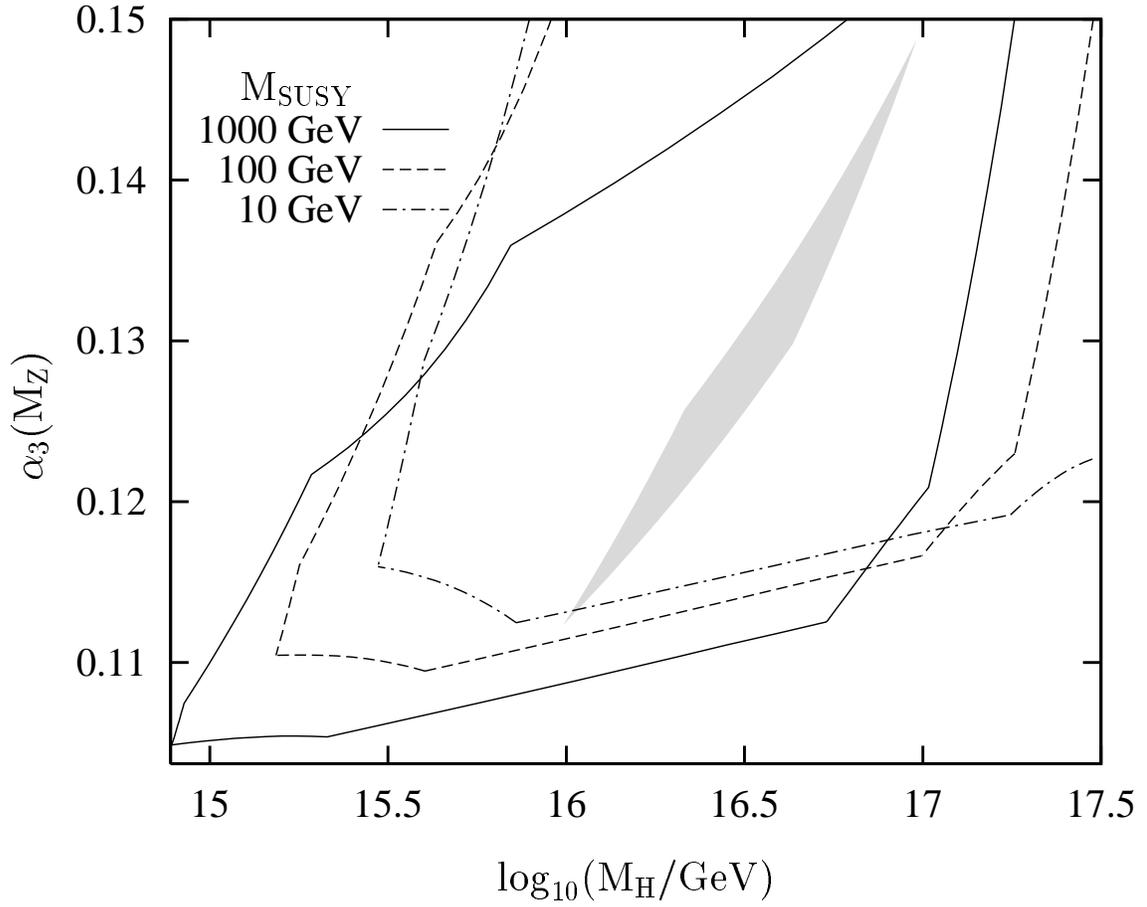

\caption{Allowed regions in the parameter space projected onto the ${\rm
M}_{\rm H}$--$\alpha_3({\rm M}_{\rm Z})$ plane.  The curves are for ${{\rm
M}_{\rm SUSY}} = 10 \ {\rm GeV}$ (solid), $100 \ {\rm GeV}$ (dashed), and $1000
\ {\rm GeV}$ (dot dashed).  The shaded region corresponds to the case where $c
= 0$, with all the GUT threshold mass scales taken to be degenerate, and ${{\rm
M}_{\rm SUSY}}$ allowed to vary between $10 \ {\rm GeV}$ to $1000 \ {\rm GeV}$.
}
\label{fig1}
\end{figure}

In order to provide a bound relating $\alpha_3$ to ${\rm M}_{\rm H}$ and ${\rm
M}_i$ we fix $c$ and $\sin^2(\theta_W)$ to take their maximum values
\cite{foot6} and $\lambda_1$ to take its minimum value, while $\lambda_2$
varies. At each point we iterate numerically in order to find the solution of
Eq.~(\ref{eq_main}) and thus $\alpha_3$ and ${\rm M}_{\rm H}$.  Some of the
results are displayed in Fig.~\ref{fig1},
%\figa
where we have chosen the degenerate case of ${\rm M}_1 = {\rm M}_2 = {\rm M}_3
= {{\rm M}_{\rm SUSY}}$, and we have plotted a bound each for ${{\rm M}_{\rm
SUSY}} = 10 \ {\rm GeV}$, $100 \ {\rm GeV}$, and $1000 \ {\rm GeV}$.

The relevant bound on $\alpha_3$, the lower horizontal line in Fig.~\ref{fig1},
can be parametrized by \cite{foot10}
\begin{eqnarray}
\alpha_{3,min}  &=& 0.040 + 0.0139 \; t_{\rm H}  \nonumber \\
		&&- 0.00579 \; t_{\rm SUSY} - 0.00454 \; t_{\rm SUSY}^2,
\label{eq_a3min}
\end{eqnarray}
where $t_\alpha \equiv (1/2 \pi) \ln ({\rm M}_\alpha/{\rm M}_{\rm Z})$ for mass
${\rm M}_\alpha$. For the case that the ${\rm M}_i$'s are non-degenerate, we
may still use Eq.~(\ref{eq_a3min}) to a close approximation, when we replace
$t_{\rm SUSY}$ using the following formula:
\begin{eqnarray}
t_{\rm SUSY} &=& 0.305 \; t_1 + 7.738 \; t_2 - 7.043 \; t_3 \nonumber \\
	&&+ 2.38 \; t_1^2 + 20.91 \; t_2^2 + 6.90 \; t_3^2 \nonumber \\
	&&- 20.21 \; t_1 t_2 + 12.54 \; t_1 t_3 - 22.53 \; t_2 t_3.
\label{eq_tsusy}
\end{eqnarray}
We find it useful to think of ${{\rm M}_{\rm SUSY}}$ as defined in this formula
as an effective mass scale to account for the SUSY thresholds in
Eq.~(\ref{eq_a3min}).  Thus, in Fig.~\ref{fig1}, although we have only plotted
the results for the degenerate SUSY thresholds case, we may think of the bottom
curves (which give the $\alpha_3$ bound) as valid for all ${\rm M}_i$ with
corresponding ${{\rm M}_{\rm SUSY}}$ as given by
Eq.~(\ref{eq_tsusy})\cite{foot11}.

Current bounds on the $p \rightarrow \bar \nu + K^+$ decay mode
\cite{plifetime} imply ${\rm M}_{\rm H} > 1.2 \times 10^{16} \ {\rm
GeV}$\cite{mHbound}. Using ${{\rm M}_{\rm SUSY}} = 100 \ {\rm GeV}$, (a
characteristic value consistent with proton decay data), one finds $\alpha_3 >
0.112$.  Thus resolving the $\alpha_3$ measurements is a crucial test of the
model.

We also plot in Fig.~\ref{fig1} the allowed region (shaded) where $c = 0$, and
all GUT thresholds are taken to be degenerate (i.e. ${\rm M}_{\rm H} = {\rm
M}_{\rm V} = {\rm M}_\Sigma = {\rm M}_{\rm U}$), for the case of degenerate
SUSY thresholds, and $ {\rm M}_i$ between $10 \ {\rm GeV}$ to $1000 \ {\rm
GeV}$. We see that the lower bound on ${\rm M}_{\rm H}$ implies now $\alpha_3 >
0.1145$, and there is a strong correlation between $\alpha_3({\rm M}_{\rm Z})$
and ${\rm M}_{\rm H}$.

We turn now to the Planck scale parameter $c$. Due to a cancellation, fixing
$\sin^2(\theta_W)$, $\alpha_3({\rm M}_{\rm Z})$, and the ${\rm M}_i$'s uniquely
determines ${\rm M}_{\rm H}$ when $c=0$ \cite{foot9}. Assuming in the future we
have a complete picture of the low energy physics, including the SUSY spectrum
and the proton decay rate in both major modes, we can compare this value for
$c=0$ (${\rm M}_{\rm H}(0)$) with the correct ${\rm M}_{\rm H}$ as determined
from proton decay, and therefore determine $c$. The result can be summarized in
the formula
\begin{equation}
c = \frac{\alpha_5 {{\rm M}_{\rm Pl}}}{10 \pi {\rm M}}
	\ln{{\rm M}_{\rm H} \over {\rm M}_{\rm H}(0)}.
\label{eq_c}
\end{equation}
${\rm M}$ could be determined by the $ p \rightarrow e^+ + \pi^0$ decay mode,
since this decay is governed by ${\rm M}_{\rm V} = 5 \sqrt{8 \pi \alpha_5} \;
{\rm M}$.  One may obtain $\alpha_5$ from Eq.~(\ref{eq_main}).  An approximate
value is $\alpha_5 = 1/23$. Note that a rough determination of $c$ is possible
even without the pion decay mode, e.g.\ by taking $\lambda_2 = 1$ so that ${\rm
M} = {\rm M}_{\rm H}/5$.  In this approximation, the front factor in
Eq.~(\ref{eq_c}) is $ \approx 0.3$, and while the ratio ${\rm M}_{\rm H} / {\rm
M}_{\rm H}(0)$ can vary substantially for reasonable range of low energy
physics data, owing to the logarithmic dependence, $c$ determined this way is
in fact $O(1)$ and not expected to be anomalously large.

\begin{figure}
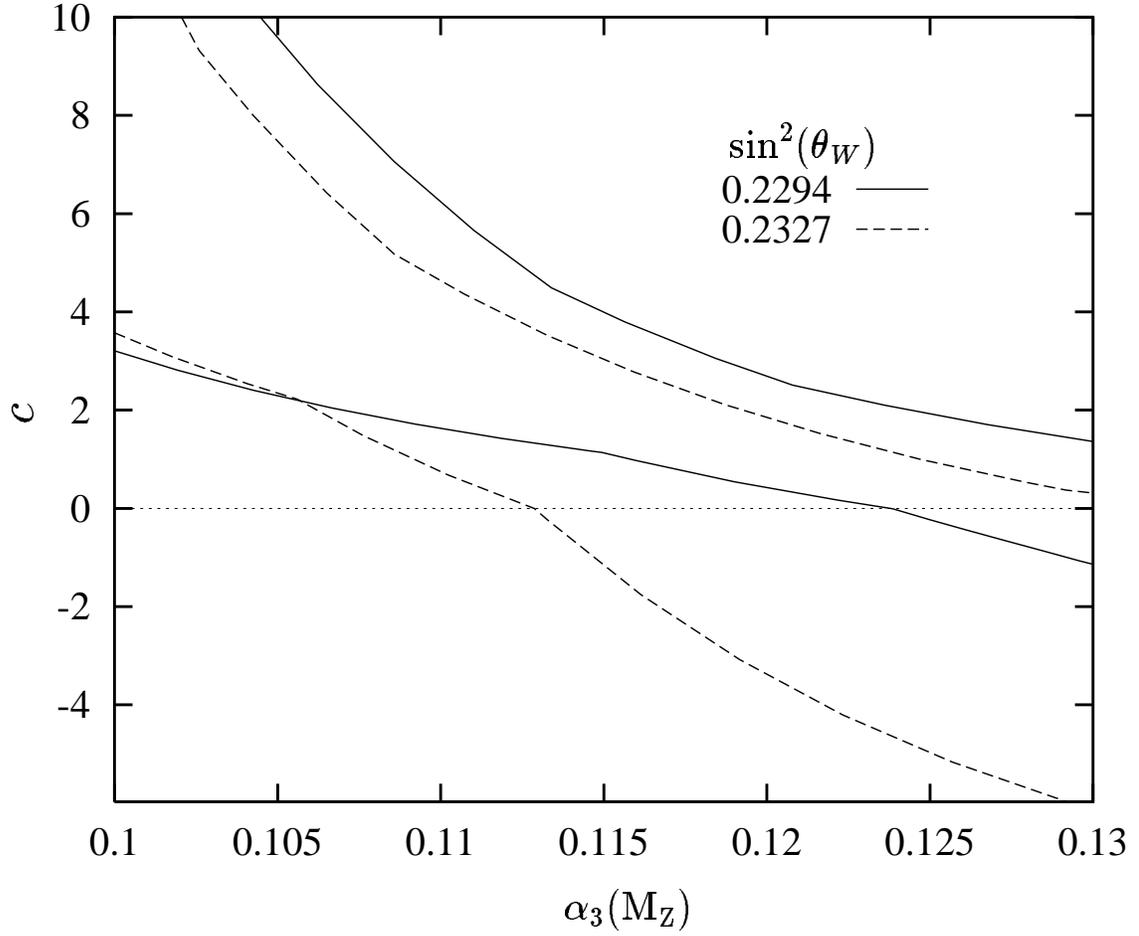

\caption{Allowed regions in the parameter space projected onto the
$\alpha_3({\rm M}_{\rm Z})$--$c$ plane.  The curves are for $\sin^2(\theta_W) =
0.2294$ (solid) and $\sin^2(\theta_W) = 0.2327$ (dashed).}
\label{fig2}
\end{figure}

Although we have not yet observed proton decay, nor been able to determine the
SUSY spectrum, using our naturalness conditions that $0.1 \le \lambda_{1,2} \le
2.0$ and $10 \ {\rm GeV} \le {\rm M}_{1,2,3} \le 1000 \ {\rm GeV}$, and
requiring ${\rm M}_{\rm H} \ge 1.2 \times 10^{16} \ {\rm GeV}$\cite{mHbound},
we can already place bounds on $c$ for different values of $\alpha_3({\rm
M}_{\rm Z})$ and $\sin^2(\theta_W)$. We plot in Fig.~\ref{fig2}
%\figb
the allowed region for $\sin^2(\theta_W) = 0.2294$ and for $\sin^2(\theta_W) =
0.2327$ in the $c$--$\alpha_3({\rm M}_{\rm Z})$ plane for ${\rm M}_1 = {\rm
M}_2 = {\rm M}_3$.  We find that in this analysis where ${\rm M}_i$'s are taken
to be degenerate, the Planck scale term must exist for some values of
$\sin^2(\theta_W)$ and $\alpha_3({\rm M}_{\rm Z})$.  We further note that
ranging over the values of $\sin^2(\theta_W)$ between 0.2294 and 0.2327, the
LEP range for $\alpha_3({\rm M}_{\rm Z})$, i.e.\ $0.117 \le \alpha_3 \le 0.129$
corresponds to $-6.0 \le c \le 3.4$; and the low energy data\cite{foot1} range,
$0.107 \le \alpha_3 \le 0.117$, corresponds to $-2.2 \le c \le 8.1$.

In contrast to the conclusions of
Refs.~\cite{barbieri/hall,faraggi,hall/sarid}, we have found that GUT models of
the type considered here do have significant experimental consequences.  Thus
for models without Planck scale terms ($c=0$), measurements made purely at the
low energy (electroweak) scale can allow a prediction of the proton lifetime,
and thus allow a direct test of the model.  When $c$ is left arbitrary, this is
no longer possible.  However, then low energy measurements will allow an
experimental determination of $c$, and one has the remarkable possibility of
seeing experimentally, for the first time, whether Planck scale physics exists.
We note further, that the presence of GUT threshold effects and Planck scale
terms do not qualitatively change the grand unification conclusions of
Ref.~\cite{amaldi_etc}.

This work was supported in part by the National Science Foundation under grant
number PHY-9411543.

\begin{references}

\bibitem[\ast]{dave} Email: Cdude@phys.tamu.edu (internet).

\bibitem[\dag]{shinichi} Email: Urano@phys.tamu.edu (internet).

\bibitem[\ddag]{dick} Email: Arnowitt@phys.tamu.edu (internet).

\bibitem{arnowitt} See, for example, R.~Arnowitt, P.~Nath, Proc VII
J.~A.~Swieca Summer School, 1993 (World Scientific, Singapore, 1994).

\bibitem{amaldi_etc} P.~Langacker, Proc.\ PASCOS 90-Symposium, Eds.\ P.~Nath
and S.~Reucroft (World Scientific, Singapore 1990); J.~Ellis, S.~Kelley and
D.~V.~Nanopoulos, Phys.\ Lett.\ {\bf B249}, 441 (1990); {\bf B260}, 131 (1991);
U.~Amaldi, W.~de Boer, H.~F\"urstenau Phys.\ Lett.\ {\bf B260} 447 (1991);
F.~Anselmo, L.~Cifarelli, A.~Peterman and A.~Zichichi, Nuov.\ Cim.\ {\bf 104A},
1817 (1991); {\bf 115A}, 581 (1992).

\bibitem{barbieri/hall} R.~Barbieri and L.~J.~Hall, Phys.\ Rev.\ Lett.\ {\bf
68}, 752 (1992).

\bibitem{faraggi} A.~Faraggi, B.~Grinstein, and S.~Meshkov, Phys.\ Rev.\ {\bf
D47} 5018 (1993)

\bibitem{hall/sarid} L.~J.~Hall and U.~Sarid, Phys.\ Rev.\ Lett.\ {\bf 70},
2673 (1993).

\bibitem{langacker/polonsky} P.~Langacker, N.~Polonsky, Phys.\ Rev.\ {\bf D47}
4028 (1993)

\bibitem{foot1}  See for example, S.~Bethke, Proc.\ XXVI Int.\ Conf.\ on High
Energy Physics, AIP Conf.\ Proc.\ No.~272 (1993) where the value $\alpha_3({\rm
M}_{\rm Z}) = 0.112 \pm 0.005$ is obtained from deep inelastic scattering
measurements.  We note that a recent lattice gauge calculation of
$\alpha_3({\rm M}_{\rm Z})$ from the $\Upsilon$ spectrum, (C.~T.~H.~Davies et
al., hep-ph/9408328 (1994)) gives $\alpha_3({\rm M}_{\rm Z}) = 0.115 \pm
0.002$,
while the current LEP evaluation is $\alpha_3({\rm M}_{\rm Z}) = 0.123 \pm
0.006$.

\bibitem{adjoint} S.~Chaudhuri, S.~Chung, and J.~Lykken, hep-ph/9405374.

\bibitem{rulin} R.~Xiu, hep-ph/9412262.

\bibitem{su5gut}
E.~Witten, Nucl.\ Phys.\ {\bf B177}, 477 (1981); {\bf B185}, 513 (1981);
S.~Dimopoulos and H.~Georgi, Nucl.\ Phys.\ {\bf B193}, 150 (1981); N.~Sakai,
Zeit.\ f.\ Phys.\ {\bf C11}, 153 (1981).

\bibitem{su5gutmore} A.~H.~Chamseddine, R.~Arnowitt, and P.~Nath, Phys.\ Rev.\
Lett.\ {\bf 49}, 970 (1982).

%\bibitem{appliedN1Sugra} P.~Nath, R.~Arnowitt, and A.~H.~Chamseddine,
%{\it Applied N=1 Supergravity} (World Scientific, Singapore, 1984).

\bibitem{su6model} K.~Inoue, A.~Kakuto and T.~Tankano, Prog.\ Theor.\ Phys.\
{\bf 75}, 664 (1986); A.~Anselm and A.~Johansen, Phys.\ Lett.\ {\bf B200}, 331
(1988); A.~Anselm, Sov.\ Phys.\ JETP {\bf 67}, 663 (1988); R.~Barbieri,
G.~Dvali and A.~Strumia, Nucl.\ Phys.\ {\bf B391}, 487 (1993).

\bibitem{hill_etc} C.~T.~Hill, Phys.\ Lett.\ {\bf 47} 135 (1984); Q.~Shafi,
C.~Wetterlich, Phys.\ Rev.\ Lett.\ {\bf 52} 875 (1984).

\bibitem{foot2} Claims in the literature \cite{faraggi} that the GUT scale can
approach the Planck scale have relied on one-loop RGE's.  With two-loop
analysis, we find in general that ${{\rm M}_{\rm GUT}}$ stays well below ${{\rm
M}_{\rm Pl}}$.

\bibitem{foot7} The Yukawa's enter in the RGE's at the two-loop level. Although
its effect is small, we include the top Yukawa term in our calculations.  We
use $b_{it} = (2/3, 1/2, 17/30)_i$.

\bibitem{einhorn/jones} M.~B.~Einhorn and D.~R.~T.~Jones, Nucl.\ Phys.\ {\bf
B196}, 475 (1982).

\bibitem{foot3} The advantage of the $\rm \overline{DR}$ renormalization scheme
here is that the ${\rm M}_i$'s for a single particle threshold will equal the
particle's mass.  Thus the ``match and run'' method sets the ${\rm M}_i$'s to
the mass of the particle.

\bibitem{foot4}  At the GUT threshold, this condition is not met for $\beta_3$.
Consequently, we do not use such a parametrization to treat the GUT threshold.

\bibitem{foot5}  The top threshold is treated this way as well, although it has
little effect on the results.

\bibitem{cdf} CDF Collaboration (F.~Abe, et al.), Phys.\ Rev.\ Lett.\ {\bf 73},
225 (1994).

\bibitem{foot6}We take $\sin^2(\theta_W)$ to lie between 0.2294 and 0.2327 and
$\alpha({\rm M}_{\rm Z})$ to be $1 / 127.9$.  All low energy data is considered
renormalized at ${\rm M}_{\rm Z}$ in the $\overline{\rm MS}$ scheme, while the
couplings run in the $\overline{\rm DR}$ scheme. The conversion between schemes
gives a negligible contribution to the results.

\bibitem{foot10} Conceptually, the relation in Eq.~(\ref{eq_a3min}) is found
from Eq.~(\ref{eq_main}) by eliminating $\alpha_5$ and $\lambda_2$, and then
solving for $\alpha_3$ in terms of ${\rm M}_i$ and ${\rm M}_{\rm H}$ while
fixing $c$, $\lambda_1$, and $\sin^2(\theta_W)$ at their extrema values of 1,
0.1, and 0.2327, respectively. We avoid treating $\sin^2(\theta_W)$ as a free
parameter in Eq.~(\ref{eq_a3min}) in order not to further complicate the
dependences.  We note, however, that as the value of $\sin^2(\theta_W)$ is
varied to its lowest allowed value (0.2294), $\alpha_{3,min}$ may move upwards
by as much as 0.007 for some choices of ${\rm M}_i$ and ${\rm M}_{\rm H}$.

\bibitem{foot11}  Eq.~(\ref{eq_tsusy}) was gotten from a numerical fit. We have
confirmed that the fit is good for variations of ${\rm M}_i$ by as much as a
factor of 2 from ${{\rm M}_{\rm SUSY}}$. Preliminary study does indicate that a
range for ${{\rm M}_{\rm SUSY}}$ of $10$ to $1000 \ {\rm GeV}$ corresponds to a
reasonable range of low energy sparticle spectrum in this model.

\bibitem{plifetime} The experimental bound on the $p \rightarrow \bar \nu K^+ $
mode is $\tau( p \rightarrow \bar \nu K^+) > 1 \times 10^{32}$ yr (90 \% CL)
from Kamiokande.  (Particle Data Group, Phys.\ Rev.\ {\bf D50}, Part 1 (1994).)

\bibitem{mHbound} R.~Arnowitt and P.~Nath, Phys.\ Rev.\ Lett.\ {\bf 69}, 725
(1992); P.~Nath and R.~Arnowitt, Phys.\ Lett.\ {\bf B289}, 368 (1992);
R.~Arnowitt and P.~Nath, Phys.\ Rev.\ {\bf D49}, 1479 (1994).

\bibitem{foot9} This can be seen by dotting in the vector $(1,-3,2)_i$ into
Eq.~(\ref{eq_main})\cite{hall/sarid}.

\end{references}
\end{document}